\newcommand{\extended}[1]{}    
\newcommand{\short}[1]{#1}     
\renewcommand{\extended}[1]{#1} 
\renewcommand{\short}[1]{}      
\documentclass{llncs}

\usepackage{url}
\usepackage[utf8]{inputenc}
\usepackage{amsfonts,amssymb}\usepackage{graphicx}
\usepackage{mathtools}
\usepackage{longtable}
\usepackage{latexsym}
\usepackage[table]{xcolor}
\usepackage[mathscr]{eucal}
\usepackage[refpage,english]{nomencl}
\usepackage{xspace}
\usepackage{subfigure}
\usepackage{wrapfig}
\usepackage{url}
\usepackage{balance}
\usepackage{multirow}
\usepackage{multicol}
\usepackage{thmtools}
\usepackage{thm-restate}
\usepackage{hyperref}

\usepackage{tikz}
\usetikzlibrary{arrows,automata,calc,shapes,decorations,backgrounds,petri,mindmap,fit,positioning}
\newcommand{\lan}[1]{\ensuremath{\mathbf{#1}}\xspace}

\newcommand{\CSL}[1][]{\lan{CSL}}
\newcommand{\CSLP}[1][]{\lan{CSLP}}

\newcommand{\MIATL}[1][]{IATL[R]}

\newcommand{\ATLES}[1][]{\lan{ATLES}}
\newcommand{\ATELA}[1][]{\lan{ATELA}}

\newcommand{\TransArrow}[1][]{\hookrightarrow....}

\newcommand{\act}[1]{{\rm #1}}

                                    \newcommand{\onepath}[1][]{\ensuremath{\lambda\ifthenelse{\equal{#1}{}}{}{[#1]}}}

\newcommand{\plaus}[1][]{\ifthenelse{\equal{#1}{}}{\mathbf{P\;\!\!l}\,}{\mathbf{P\;\!\!l}_{#1}\,}}
\newcommand{\phys}[1][]{\ifthenelse{\equal{#1}{}}{\mathbf{P\;\!\!h}\,}{\mathbf{P\;\!\!h}_{#1}\,}}

\newcommand{\plaumodels}[1][]{\ensuremath{\ifthenelse{\equal{#1}{}}{\models_\sPlaupaths}{\models_{#1}}}}
\newcommand{\sPlaupaths}{\ensuremath{P}}

\newcommand{\true}{\mathit{true}}
\newcommand{\false}{\mathit{false}}

\definecolor{lightgrey}{rgb}{0.8,0.8,0.8}
\definecolor{grey}{rgb}{0.6,0.6,0.6}
\definecolor{darkgrey}{rgb}{0.4,0.4,0.4}
\definecolor{darkgreen}{rgb}{0,0.7,0}

\newcommand{\set}[1]{\{{#1}\}}

\newcommand{\argmax}{\mathrm{argmax}}

\newcommand{\Nat}{\mathbb{N}}

\newcommand{\putaway}[1]{}

\newcommand{\para}[1]{\smallskip\noindent\textbf{#1}}

\newenvironment{enumerate2}{\begin{enumerate}\itemsep 0in}{\end{enumerate}}

\newcommand{\finis}{{\scriptsize $\blacksquare$}}
\newcommand{\finisdef}{$\Box$}
\newcommand{\bul}{{\tiny $\blacksquare$}}

\def\itemiremember{\labelitemi}
\def\itemiiremember{\labelitemii}

\newcommand{\opp}[1]{{\color{darkgrey}#1}}

\newcommand{\nash}[1]{\framebox{{#1}}}
\newcommand{\maxmin}[1]{\mathbf{{#1}}}
\newcommand{\BR}[1]{\underline{{#1}}}
\newcommand{\stackelberg}[1]{\colorbox{yellow}{{#1}}}

\newcommand{\NatOne}{\Nat_{\ge 1}}

\newcommand{\nmax}{{n_{max}}}
\newcommand{\nfalse}{n_{\mathit{cheat}}}
\newcommand{\ncast}{n_{cast}}
\newcommand{\caudit}{c_{audit}}
\newcommand{\asucc}[1]{\mathit{Succ}_{#1}}
\newcommand{\afail}[1]{\mathit{Fail}_{#1}}
\newcommand{\acatch}[1]{\mathit{Catch}_{#1}}
\newcommand{\vsucc}{\asucc{V}}
\newcommand{\vfail}{\afail{V}}
\newcommand{\vcatch}{\acatch{V}}
\newcommand{\dsucc}{\asucc{D}}
\newcommand{\dfail}{\afail{D}}
\newcommand{\devcolor}[1]{{#1}}

\newcommand{\strat}{s}

\renewcommand{\act}{\alpha}
\newcommand{\mathNE}{{_\mathit{NE}}}
\newcommand{\StackVal}{\mathit{S\!Val}_V}
\newcommand{\payvect}[1]{\,\fbox{\ensuremath{#1}}\,}
\newcommand{\payvectout}[1]{\,\fbox{\ensuremath{#1}}\,}

\newcommand{\WJ}[1]{{\color{darkgreen}[XX: #1]}}
\newcommand{\todo}[1]{{\color{red}[TO DO: #1]}}
\renewcommand{\WJ}[1]{}
\renewcommand{\todo}[1]{}

\title{Pretty Good Strategies for Benaloh Challenge}
\author{Wojciech Jamroga}
\institute{
  Interdisc. Centre on Security, Reliability and Trust, SnT, University of Luxembourg \\
  Institute of Computer Science, Polish Academy of Sciences, Warsaw, Poland \\
  \email{wojciech.jamroga@uni.lu} }

\begin{document}
\pagestyle{plain}
\maketitle

\begin{abstract}
Benaloh challenge allows the voter to audit the encryption of her vote, and in particular to check whether the vote has been represented correctly. An interesting analysis of the mechanism has been presented by Culnane and Teague. The authors propose a natural game-theoretic model of the interaction between the voter and a corrupt, malicious encryption device. Then, they claim that there is no ``natural'' rational strategy for the voter to play the game. In consequence, the authorities cannot provide the voter with a sensible auditing strategy, which undermines the whole idea.

Here, we claim the contrary, i.e., that there exist simple rational strategies that justify the usefulness of Benaloh challenge.
\end{abstract}

\WJ{Extended version:

- Where to submit:
Social Choice and Welfare (70), Games and Economic Behavior (140),  ACM Transactions on Economics and Computation (70) ? \\
Journal of Information Security and Applications (100; IF 4.96; topic: Human factors in security), ACM Transactions on Information and System Security (140) -> was transformed to ACM Transactions on Privacy and Security (20)

- What to add:
detailed proofs, more discussion of the model (esp payoffs), more detailed GT primer, characterization of SV for $\nmax > 2$ ?
}

\todo{Address PYAR's comments:
\begin{enumerate}
\item If D has a notion how V will cast, say $c_0$, and V inputs $c_1$ then it’s a good guess that V will audit, so don’t cheat, especially if $c_1$ is an unpopular candidate . Conversely if V inputs $c_0$ then will probably cast, especially if this is the second try after inputing something else on the first try, so cheat.

    \underline{Answer}: Vanessa and Chris argued in their paper that it doesn't pay off for the voter to encrypt a vote different from the voter's true preferences, and I agree with their argument. (BTW, their reasoning is very similar to what you wrote.) So, I assumed that the voter always submits her true vote for encryption, even if she intends to audit.
    Well spotted - this is by no means obvious, and should be stated explicitly!

\item Also I’m curious as to the intuition as to why Stackelberg gives V a better expected payoff than Nash.

    \underline{Answer}: Stackelberg is always a bit of a tradeoff. On the one hand the leader takes the initiative and puts the follower on the defensive. On the other hand, the leader gives away valuable information that the follower can exploit.
    Interestingly, the former often outweighs the latter, which seems counterintuitive. It might be because our basic intuition comes from zero-sum games, and there being the leader is always detrimental. But in general-sum games choosing the "battle ground" is often more important than exposing your strategy to the enemy.

\item A lot of the reasoning here, and in T+C, is based on the assumption that being caught once or at least a small number of times will be catastrophic for the attacker. In reality this probably isn’t really true due to difficulties with dispute resolution, device faking crashing when audited etc.

    \underline{Answer}: Well, maybe not catastrophic, but I'd say getting caught brings penalty that is disproportionately large compared to the reward for swinging one vote. If you swing one vote, you gain one vote. If you get caught on swinging one vote, there is a non-negligible probability that you will lose *most of the affected votes*. And it increases with every next vote on which you are caught tampering.

\item Being caught just once across an election presumably won’t trigger much, only a number of indigents, maybe with some pattern would trigger investigation etc. but this just influences the payoff. But I would guess that an attacker would be ok with a strategy that has an expectation of a small number of detections.

    \underline{Answer}: True. But by the same token swinging one vote is also of little use to the attacker. In this sense, both the attacker's reward (for successfully cheating) and the penalty (for getting caught) is nonlinear wrt the number of successful/detected swaps.
    We use an idealized model where both the reward and the penalty of the attacker is additive, and thus can be decomposed into 1-1 interactions between a single voter and the attacker-controlled device.
\end{enumerate}
}

\todo{
Address reviewers' comments:

\begin{enumerate}
\item Reviewer 2: One interesting question on this, the analysis was performed in isolation – as a direct game between the voter and the device; it will be interesting to see how this compares when the adversary controls multiple devices and plays multiple rounds with different voters. I think such an adversary will have a stronger chance of winning then just running many independent 1-to-1 sessions with voters. This would also allow to capture more realistic adversaries as they can control multiple casting devices [if the voter uses their own pc to cast ballots].

\item Reviewer 3: It would be interesting to see a more thorough discussion on the payoffs as consequences of either failing or succeeding for either the voter or the device. In particular, the paper assumes that the costs of failure for the device are higher than the benefits of success; I wonder, however, whether this is the case in practice. For once, I imagine it depends on whether the voter reports the failure in the first place (from studies we know that at least some voters would just assume that it was them who made a mistake, or that the failure is a result of a system bug rather than a malicious attack) or just disregard the attack and vote via a different channel. Then even in case the voter decides to report, the actual consequences to the attacker would depend on the actions taken by the election authorities, and whether they would follow up on the report at all; to which extent these actions are being/will be taken can be seen as a discussion topic by itself.
    $\leadsto$ Peter's comments
\end{enumerate}
}

\section{Introduction}\label{sec:intro}

\emph{Benaloh challenge}~\cite{Benaloh06verifiable,Benaloh07challenge} aims to give the voter the possibility to audit the encryption of her vote, and in particular to check whether the vote has been represented correctly.
More precisely, the device that encrypts and sends the ballot must first commit to a representation of the vote\extended{ given as input}. After that, the voter decides whether to cast it or ``spoil'' it, i.e., open the encryption and check its correctness.
Intuitively, this should reduce the risk of altering the value of the vote by a malfunctioning or corrupt machine when it casts the ballot on the voter's behalf.

An interesting analysis of the mechanism has been presented in~\cite{Culnane16benalohGT}. The authors propose a natural game-theoretic model of the interaction between the voter and a corrupt, malicious encryption device. Then, they claim that there is no ``natural'' rational strategy for the voter to play the game\extended{, where rational play is defined in terms of Nash equilibrium~\cite{Nash50equilibrium}}. More precisely, they claim that: (1) only randomized voting strategies can form a Nash equilibrium, (2) for audit sequences with bounded length, the voter gets cheated in all Nash equilibria, and (3) the Nash equilibria in the infinite game do not form an easy pattern (e.g., Bernoulli trials).
In consequence, the voter cannot be provided with a sensible auditing strategy, which undermines the whole method.

In this paper, we claim that -- on the contrary -- there exist simple auditing strategies that justify the usefulness of Benaloh challenge. This follows from three important observations.
First, we show that there \emph{are} Nash equilibria in bounded strategies where the voter casts her intended vote with high probability.
Based on this observation, we focus on a small subset of randomized strategies, namely the ones where the voter spoils the ballot with probability $p$ in the first round, and in the second round always casts.
Secondly, we point out that the rationality of strategies in Benaloh challenge is better captured by Stackelberg equilibrium\extended{~\cite{Stackelberg34equilibrium,Stackelberg52market,Leitmann78stackelberg}}, rather than Nash equilibrium.
Thirdly, a sensible Stackelberg strategy does not have to be optimal; it suffices that it is ``good enough'' for whatever purpose it serves.
Fourthly, we prove that the\extended{ generalized} Stackelberg equilibrium in the set of such strategies does not exist, but the voter can get arbitrarily close to the upper limit of the Stackelberg payoff. To show this, we formally define the concept of \emph{Stackelberg value}, and show that it is always higher than the value of Nash equilibrium in the set of randomized strategies for the voter.

\para{Related work.}
Game-theoretic analysis of voting procedures that takes into account the economic or social incentives of the participants has been scarce.
In~\cite{Buldas07evoting}, two voting systems were compared using zero-sum two-player games based on attack trees, with the payoffs representing the success of coercion.
In~\cite{Jamroga17preventing}, a simple game-theoretic model of preventing coercion was proposed and analyzed using Nash equilibrium, maxmin, and Stackelberg equilibrium.
The authors of~\cite{Yin16protectingelections} applied Stackelberg games to prevent manipulation of elections, focussing on the computational complexity of preventing Denial of Service attacks.
The research on \emph{security games}~\short{\cite{Tambe11securitygames}}\extended{\cite{Yin10stackelberg,Tambe11securitygames,Fang15anti-poaching}}, using Stackelberg equilibrium to design anti-terrorist and anti-poaching policies, is of some relevance, too.

\section{Benaloh Challenge and Benaloh Games}\label{sec:benaloh}

We start by a brief introduction of Benaloh challenge. Then, we summarize the game-theoretic analysis of the challenge, proposed in~\cite{Culnane16benalohGT}.

\subsection{Benaloh Challenge}\label{sec:benaloh-challenge}

\emph{Benaloh challenge}~\cite{Benaloh06verifiable,Benaloh07challenge} is a ``cut-and-choose'' technique for voter-initiated encryption audits, which proceeds as follows:
\begin{enumerate}
\item\label{it:ballot-generate} An empty ballot is generated and provided to the voter.

\item The voter fills in the ballot and transmits it to the encryption device;

\item The device encrypts the ballot with the election public key, and makes the encrypted vote available to the voter;

\item The voter decides to cast the encrypted vote, or to open and audit the encryption. If the encryption is opened, the ballot is discarded, and the voter proceeds back to step~\ref{it:ballot-generate}.
\end{enumerate}

Benaloh challenge is meant to counter the threat of a malicious encryption device that falsely encrypts the ballot, e.g., in favor of another election candidate.
Importantly, this should be done without compromising receipt-freeness of the voting protocol.
In a broader perspective, the challenge can be applied in any communication scenario where the encryption mechanism is not trustworthy and plausible deniability is required on the side of the sender.

The idea behind the technique is that, if the voters audit the encryptions from time to time, corrupt devices will be exposed and investigated.
Thus, it does not pay off to tamper with the encryption in the long run, and the perpetrator would have little incentive to do that.
At its core, this is a game-theoretic argument.

\begin{figure}[t]
\resizebox{\textwidth}{!}{\begin{tabular}{|c|c|c|c|}
\hline
Condition & \cellcolor{yellow!20} Voter payoff     & \cellcolor{yellow!20} Device payoff                         & Comment \\
 & \cellcolor{yellow!20} $u_V(\ncast,\nfalse)$ & \cellcolor{yellow!20} $u_D(\ncast,\nfalse)$                &
\\ \hline
\cellcolor{yellow!20} $\ncast < \nfalse$  & $\vsucc - (\ncast-1)\caudit$ & $0$        & Voter votes as intended
\\ \hline
\cellcolor{yellow!20} $\ncast = \nfalse$  & $-\vfail - (\ncast-1)\caudit$ & $\dsucc$   & Device successfully cheats
\\ \hline
\cellcolor{yellow!20} $\ncast > \nfalse$  & $- \nfalse\cdot\caudit$ & $-\dfail$         & Voter catches cheating device
\\ \hline
\end{tabular}
}
\caption{Inspection game for Benaloh challenge~\cite[Fig.~2]{Culnane16benalohGT}}
\label{fig:Benaloh-game-NF}
\end{figure}

\subsection{Benaloh Challenge as Inspection Game}\label{sec:benaloh-game}

Intuitively, the interaction in Benaloh challenge can be seen as a game between the voter $V$ and the encryption device $D$ -- or, more accurately, between the voter and the malicious party that might have tampered with the device. We will use the term \emph{Benaloh game} to refer to this aspect of Benaloh challenge.
In each round, the voter can choose between casting her intended vote (action $cast$) and auditing the encryption (action $audit$). At the same time, the device chooses to either encrypt the vote truthfully (action $true$) or cheat and encrypt another value of the vote (action $\false$).
Both players know exactly what happened in the previous rounds, but they decide what to do without knowing what the other player has selected in the current round.

A very interesting analysis has been presented by Chris Culnane and Vanessa Teague in~\cite{Culnane16benalohGT}. The authors model the interaction as an \emph{inspection game}\extended{, i.e., a non-cooperative game where one player verifies if the other party adheres to a given requirement -- typically, a legal rule}~\cite{Avenhaus00inspectiongames}.
The idea is very simple: $V$ chooses the round $\ncast$ in which she wants to cast the vote, and $D$ chooses the round $\nfalse$ when it will fake the encryption for the first time.
Consequently, the voter's plan is to audit the encryption in all rounds $n<\ncast$, and similarly the device encrypts truthfully for all $n<\nfalse$.
The players choose their strategies before the game, without knowing the opponent's choice.
Their payoffs (a.k.a. utilities) are presented in Figure~\ref{fig:Benaloh-game-NF}, with the parameters interpreted as follows:
\begin{itemize}
\item $\asucc{i}$: the reward of player $i$ for succeeding with their task (i.e., casting the vote as intended for $V$, and manipulating the vote for $D$);
\item $\afail{i}$: player $i$'s penalty for failing (i.e., getting cheated for $V$, and getting caught with cheating for $D$);
\item $\caudit$: the cost of a single audit; essentially, a measure of effort and time that $V$ needs to invest into encrypting and spoiling a spurious ballot;
\end{itemize}
It is assumed that $\asucc{i},\afail{i},\caudit > 0$. Also, $\caudit < \vfail$, i.e., the voter cares about what happens with her vote enough to audit at least once.

There are two variants of the game: finite, where the number of rounds is bounded by a predefined number $\nmax\in\NatOne$, and infinite, where the game can proceed forever.
In the finite variant, the voter chooses $\ncast\in\set{1,\dots,\nmax}$, and the device selects $\nfalse\in\set{1,\dots,\nmax,\infty}$, with $\nfalse = \infty$ meaning that it always encrypts truthfully and never cheats. In the infinite variant, the voter and the device choose respectively $\ncast\in\NatOne$ and $\nfalse\in\NatOne\cup\set{\infty}$.
The structure of the game is common knowledge among the players.

\para{Discussion.}
One might consider a slightly richer game by allowing the voter to refuse participation ($\ncast=0$) or to keep auditing forever ($\ncast=\infty$). Also, we could include a reward $\vcatch$ that the voter gets when detecting an attack and reporting it to the authorities. In this paper, we stick to the game model of~\cite{Culnane16benalohGT}, and leave a proper analysis of the richer game for the future.

\subsection{Are There Simple Rational Strategies to Cast and Audit?}\label{sec:CT-claims}

Culnane and Teague make the following claims about their model (and, by implication, about the game-theoretic properties of Benaloh challenge):

\begin{enumerate}
\item\label{it:Nash-det}
  There is no Nash equilibrium in {deterministic strategies}~\cite[Lemma~1]{Culnane16benalohGT}. Thus, a rational voter must use \emph{randomized strategies} in Benaloh challenge.\footnote{
    A concise explanation of game-theoretic terms is presented in Sections~\ref{sec:GT-primer-short-1} and~\ref{sec:GT-primer-short-2}. }

\smallskip
\item\label{it:Nash-finite}
  A Nash equilibrium in the \emph{finite Benaloh game} can only consist of the voter casting right away and the device cheating right away; the argument proceeds by backward induction~\cite[Lemma~2 and its proof]{Culnane16benalohGT}. Thus, by~\cite[Lemma~1]{Culnane16benalohGT}, there are no Nash equilibria in the finite Benaloh game, and a rational voter should use \emph{infinite audit strategies}.

\smallskip
\item\label{it:Nash-infinite}
  In the \emph{infinite Benaloh game}, there is no Nash equilibrium in which the voter executes a Bernoulli process, i.e., randomizes in each round with the same probability $r$ whether to audit or cast~\cite[Theorem~2]{Culnane16benalohGT}.
  Quoting the authors, ``this prevents authorities from providing voters with a sensible auditing strategy.''
  In other words, there are no ``easy to use'' rational strategies for the voter in Benaloh challenge.
\end{enumerate}

The above claims have two controversial aspects: a technical one and a conceptual one.
First, while claims~(\ref{it:Nash-det}) and~(\ref{it:Nash-infinite}) are correct, claim~(\ref{it:Nash-finite}) is not.
By Nash's theorem~\cite{Nash50equilibrium}, every finite game has a Nash equilibrium in randomized strategies, and this one cannot be an exception.
We look closer at the issue in Section~\ref{sec:Benaloh-Nash}, show why backward induction does \emph{not} work here, and demonstrate that a clever election authority can design the procedure so that the voters do have a simple Nash equilibrium strategy to cast and audit.

Secondly, the authors of~\cite{Culnane16benalohGT} implicitly assume that ``sensible strategies'' equals ``simple Nash equilibrium strategies.''
As we discuss in Section~\ref{sec:Benaloh-Stackelberg}, Nash equilibrium is not the only concept of rationality that can be applied here.
In fact, Stackelberg equilibrium~\short{\cite{Stackelberg52market}}\extended{\cite{Stackelberg34equilibrium,Stackelberg52market}} is arguably a better fit for the analysis of Benaloh challenge.
Following the observation, we prove that generalized Stackelberg equilibrium~\cite{Leitmann78stackelberg} for the voter in the set of randomized strategies does not exist, but $V$ can get arbitrarily close to the upper limit of the Stackelberg payoff function. Moreover, there is always a Bernoulli strategy for the voter whose Stackelberg value is higher than the payoff in Nash equilibrium.
In sum, Stackelberg games better capture rational interaction in Benaloh challenge, provide the voter with simple strategies, and obtain higher payoffs for $V$ than Nash equilibria.

\begin{figure}[t]\centering
  \begin{minipage}{0.42\textwidth}\centering
    \begin{tabular}{c@{\ }|@{\ }c@{\quad}c}
    $Alice\ \backslash\ \opp{Bob}$ & \opp{\textit{bar}} & \opp{\textit{theater}} \\ \hline
    \textit{bar} & \stackelberg{$3,\opp{\BR{2}}$} & $1,\opp{0}$\putaway{$\maxmin{1,\opp{0}}$} \\
    \textit{theater} & $\BR{4},\opp{0}$ & \nash{$\BR{2},\opp{\BR{3}}$}
    \end{tabular}

    \smallskip
    \caption{A variation on the Battle of the Sexes game. The only Nash equilibrium is indicated by the black frame. \putaway{Maxmin for Alice is highlighted in bold and }Stackelberg equilibrium for Alice is set on yellow background. The players' best responses to the opponent's strategies are underlined }
    \label{fig:BoS-NFgame}
  \end{minipage}
\hfill
\begin{minipage}{0.52\textwidth}\centering
    \begin{tikzpicture}[->,>=stealth',shorten >=1pt,auto,node distance=2.1cm,transform shape,semithick,scale=0.9]
    ﻿\tikzstyle{state}=[circle,fill=none,draw=black,text=black,font=\footnotesize] \tikzstyle{initstate}=[state,fill=yellow] \tikzstyle{finalstate}=[state,fill=black] \tikzstyle{trans}=[font=\footnotesize,anchor=center]

\node[initstate] (s0) {};
\node[finalstate] (s1) [below left=1.5cm and 1.5cm of s0,label=below:{$1, \opp{1}$}] {};
\node[state] (s2) [below right=1.5cm and 1.5cm of s0] {};
\node[finalstate] (s6) [below left=1.5cm and 2.1cm of s2,label=below:{$3,\opp{2}$}] {};
\node[finalstate] (s7) [below left=1.5cm and 0.7cm of s2,label=below:{$1,\opp{0}$}] {};
\node[finalstate] (s8) [below right=1.5cm and 0.7cm of s2,label=below:{$4,\opp{0}$}] {};
\node[finalstate] (s9) [below right=1.5cm and 2.1cm of s2,label=below:{$2,\opp{3}$}] {};

\path
(s0)
  edge node[trans] {${\large\substack{stay,stay \\ out,stay \\ stay,out}}$} (s1)
  edge node[trans] {$out,out$} (s2)
(s2)
  edge[grey] node[trans,sloped,above] {$bar,bar$} (s6)
  edge[grey] node[trans] {$bar,th$} (s7)
  edge node[trans,below] {$th,bar$} (s8)
  edge node[trans,sloped,above] {$th,th$} (s9)
  ;
     \end{tikzpicture}
\caption{Multi-step Battle of the Sexes. The initial state is filled with yellow, and terminal states with black. Transitions corresponding to dominated choices are shown in grey}
    \label{fig:BoS-EFgame}
  \end{minipage}
\end{figure}

\section{Intermezzo: Game Theory Primer, Part One}\label{sec:GT-primer-short-1}

Here, we present a compressed summary of the relevant game-theoretic notions.
For a detailed introduction, see e.g.~\cite{Osborne94gamet,Shoham09MAS}.

\para{Strategic games.}
A \emph{strategic game} consists of a finite set of \emph{players} (or \emph{agents}), each endowed with a finite set of \emph{actions}.
A tuple of actions, one per player, is called an \emph{action profile}. The \emph{utility  function} $u_i(\alpha_1,\dots,\alpha_n)$ specifies the \emph{utility} (often informally called the \emph{payoff}) that agent $i$ receives after action profile $(\alpha_1,\dots,\alpha_n)$ has been played.
In the simplest case, we assume that each player plays by choosing a single action. This kind of choice represents a \emph{deterministic strategy} (also called \emph{pure strategy}) on the part of the agent.

The payoff table of an example strategic game is shown in Figure~\ref{fig:BoS-NFgame}. Two players, Alice and Bob, decide in parallel whether to go to the local bar or to the theater. The strategies and utilities of Bob are set in grey for better readability.

\para{Rationality assumptions.}
The way rational players choose their behaviors is captured by \emph{solution concepts}, formally represented by a subset of strategies or strategy profiles.
In particular, \emph{Nash equilibrium (NE)} selects those strategy profiles $\sigma$ which are stable under unilateral deviations, i.e., no player $i$ can improve its utility by changing its part of $\sigma$ while the other players stick to their choices.
Equivalently, $\sigma$ is a Nash equilibrium if each $\sigma_i$ is a best response to the choices of the other players in $\sigma$.
In our example, \textit{(theater,theater)} is the only Nash equilibrium.
Another solution concept (Stackelberg equilibrium) will be introduced in Section~\ref{sec:GT-primer-short-2}.

\para{Multi-step games.}
To model multi-step interaction, we use \emph{concurrent extensive form games}, i.e., game trees where the players proceed in rounds, and choose their actions simultaneously in each round.
The agents' payoffs are defined for each \emph{play}, i.e., maximal path from the root to a leaf of the tree.
A multi-step variant of the Battle of the Sexes, where Alice and Bob first veto-vote on whether to go out and then decide on where to go, is shown in Figure~\ref{fig:BoS-EFgame}.
In such games, a deterministic strategy of player $i$ is a conditional plan that maps the nodes in the tree to $i$'s actions.
Each strategy profile determines a unique play.

Nash equilibrium is defined analogously to strategic games.
Additionally, $\sigma$ is a \emph{subgame-perfect Nash equilibrium (SPNE)} if it is a Nash equilibrium in each subtree obtained by fixing another starting point for the game.
\emph{Backward induction} eliminates choices that are \emph{weakly dominated}, i.e., ones for which there is another choice obtaining a better vector of payoffs.
Backward induction preserves subgame-perfect Nash equilibria, and can be used to reduce the game tree if the agents are assumed to play SPNE.
For example, Alice's strategy \textit{bar} obtains payoff vector $\payvectout{\payvect{3}\payvect{1}}$, while \textit{theater} obtains $\payvectout{\payvect{4}\payvect{2}}$.
Thus, the former strategy is dominated by the latter, and can be removed from the game three.

\para{Randomized play.}
Randomization makes it harder for the opponents to predict the player's next action, and to exploit the prediction.
Moreover, Nash equilibrium is guaranteed to exist for randomized strategy profiles (Nash's theorem\extended{~\cite{Nash50equilibrium}})\extended{, whereas no such guarantee applies to pure strategies}.
In multi-step games, players can randomize in two ways.
A \emph{mixed strategy} for player $i$ is \extended{represented by }a probability distribution over the pure strategies of $i$, with the idea that the player randomizes according to that distribution, and then duly executes the selected multi-step strategy.
A \emph{behavioral strategy} assigns each game node with a probability distribution over the \emph{actions} of $i$, with the idea that $i$ randomizes freshly before each subsequent move.
By Kuhn's theorem, every mixed strategy has an outcome-equivalent behavioral strategy\extended{~\cite{Kuhn50extensiveGames}} and vice versa\extended{~\cite{Hart92games}} in games with perfect recall\extended{ (i.e., ones where players never forget what they have observed)}.
Note that deterministic strategies can be seen as a special kind of randomized strategies that use only Dirac distributions, i.e., $\strat_i(\act) = 1$. In that case we will write $\strat_i = \act$ as a shorthand.

\section{Benaloh According to Nash}\label{sec:Benaloh-Nash}

In this section, we look closer at the claims of~\cite{Culnane16benalohGT}.

\subsection{Deterministic Audit Strategies in Benaloh Games}\label{sec:Nash-det}
\label{sec:Nash-finite}

The first claim\extended{ of Culnane and Teague} is that Benaloh games have no Nash equilibrium where the voter plays deterministically~\cite[Lemma~1]{Culnane16benalohGT}. This is indeed true.
To see that, consider any strategy profile $(\ncast,\strat_D)$ where $V$ deterministically chooses a round $\ncast$ to cast her vote, and $D$ chooses $\nfalse$ according to probability distribution $\strat_D$.
If $\strat_D \neq \ncast$, then the device increases its payoff by responding with $\strat_D = \ncast$, i.e., cheating with probability $1$ at round $\ncast$; hence, $(\ncast,\strat_D)$ is not a Nash equilibrium.
Conversely, if $\strat_D = \ncast$, then the voter increases her payoff by changing her mind and casting at round $\ncast-1$ earlier (if $\ncast>1$) or at round $\ncast+1$ (otherwise); hence $(\ncast,\ncast)$ is not a Nash equilibrium either.

Ultimately, $V$ must use randomized strategies, so that $D$ cannot precisely predict in which round the vote will be cast.

\begin{figure}[t]\centering
  \begin{tikzpicture}[->,>=stealth',shorten >=1pt,auto,node distance=2.1cm,transform shape,semithick,scale=0.9]
  ﻿\tikzstyle{state}=[circle,fill=none,draw=black,text=black,font=\footnotesize] \tikzstyle{initstate}=[state,fill=yellow] \tikzstyle{finalstate}=[state,fill=black] \tikzstyle{trans}=[font=\footnotesize,anchor=center]

\node[initstate] (s0) {};
\node[finalstate] (s1) [below left=1.5cm and 3cm of s0,label=left:{$\substack{\vsucc\\ \\ \devcolor{0}}$}] {};
\node[finalstate] (s2) [below left=1.5cm and 1cm of s0,label=below:{$\substack{-\vfail\\ \\ \devcolor{\dsucc}}$}] {};
\node[state] (s3) [below right=1.5cm and 1cm of s0] {};
\node[finalstate] (s4) [below right=1.5cm and 3cm of s0,label=below:{$\substack{-\caudit\\ \\ \devcolor{-\dfail}}$}] {};
\node[state] (s5) [below of=s3] {};
\node[finalstate] (s6) [below left=1.5cm and 3cm of s5,label=left:{$\substack{-(\nmax-2)\caudit + \vsucc\\ \\ \devcolor{0}}$}] {};
\node[finalstate] (s7) [below left=1.5cm and 1cm of s5,label=below:{$\substack{-(\nmax-2)\caudit - \vfail\\ \\ \devcolor{\dsucc}}$}] {};
\node[state] (s8) [below right=1.5cm and 1cm of s5] {};
\node[finalstate] (s9) [below right=1.5cm and 3cm of s5,label=below:{$\substack{-(\nmax-1)\caudit\\ \\ \devcolor{-\dfail}}$}] {};
\node[finalstate] (s10) [below left=1.5cm and 1cm of s8,label=below:{$\substack{-(\nmax-1)\caudit + \vsucc\\ \\ \devcolor{0}}\qquad\qquad$}] {};
\node[finalstate] (s11) [below right=1.5cm and 1cm of s8,label=below:{$\qquad\qquad\substack{-(\nmax-1)\caudit - \vfail\\ \\ \devcolor{\dsucc}}$}] {};

\path
(s0)
  edge node[trans,sloped,above] {$cast,true$} (s1)
  edge node[trans] {$cast,\false$} (s2)
  edge node[trans,below] {$audit,true$} (s3)
  edge node[trans,sloped,above] {$audit,\false$} (s4)
(s3)
  edge[dotted] (s5)
(s5)
  edge node[trans,sloped,above] {$cast,true$} (s6)
  edge node[trans] {$cast,\false$} (s7)
  edge node[trans,below] {$audit,true$} (s8)
  edge node[trans,sloped,above] {$audit,\false$} (s9)
(s8)
  edge[grey] node[trans] {$cast,true$} (s10)
  edge node[trans,below] {$cast,\false$} (s11);
   \end{tikzpicture}
\caption{Game tree for Benaloh challenge. $V$'s payoffs are in black, $D$'s payoffs in red}
\label{fig:Benaloh-game-EF}
\end{figure}

\subsection{The Rise and Fall of Backward Induction}\label{sec:backinduction}

Now, we turn to randomized voting strategies in Benaloh games with finite horizon $\nmax$.
It was claimed in~\cite[proof of Lemma~2]{Culnane16benalohGT} that all $V$'s strategies where the voter does not cast immediately cannot be part of a Nash equilibrium.
The argument goes by backward induction: $D$ knows that $V$ must cast in round $n=\nmax$, so it can safely cheat in that round.
Thus, the voter should cast in rounds $1,\dots,\nmax-1$ to avoid being cheated, in which case the device can actually safely cheat in round $\nmax-1$, and so on. Unfortunately (or fortunately from the voters' point of view), the argument is incorrect.

To begin with, backward induction \emph{cannot} be applied to games in strategic form nor to inspection games; it requires a proper representation of the sequential nature of the game.
We propose the concurrent EF game in Figure~\ref{fig:Benaloh-game-EF} as a model of Benaloh challenge with horizon $\nmax$.
Each level in the game tree corresponds to a subsequent round of the game. The players choose their actions simultaneously; if $V$ casts, or $V$ audits and $D$ submits false encryption, then the game ends and the payoffs are distributed. If $V$ audits and $D$ encrypts truthfully, the game proceeds to the next round. At $n=\nmax$, the voter can only cast.

Let us start with the final round of the procedure (i.e., the lowest level in the tree). $D$ has two available choices: $\true$ and $\false$, promising the payoff vectors of $\payvect{0}$ and $\payvect{\asucc{D}}$, respectively. Indeed, the choice to encrypt truthfully is dominated and can be removed from the tree, leaving only the right-hand branch. We can also propagate the payoffs from the remaining leaf to its parent (i.e., $-(\nmax-1)\caudit - \vfail$ for $V$, and $\dsucc$ for $D$).

Consider now the second-to-last level of the tree. Again, the device has two choices: $true$ promising $\payvectout{\payvect{0}\payvect{\dsucc}}$, and $\false$ promising $\payvectout{\payvect{\dsucc}\payvect{-\dfail}}$. It is easy to see that none of them dominates the other: $\false$ works strictly better if the opponent decides to cast, whereas $true$ obtains better payoff if the opponent does $audit$.
Also the voter has now two available choices: $cast$ with the payoff vector $\payvectout{\payvect{-(\nmax-2)\caudit + \vsucc}\payvect{-(\nmax-2)\caudit - \vfail}}$ and $audit$ with $\payvectout{\payvect{-(\nmax-1)\caudit - \vfail}\payvect{-(\nmax-1)\caudit}}$. Clearly, the former vector obtains better payoff in the first dimension, but strictly worse in the second one. Thus, no choice of the voter is dominated.
Since we cannot eliminate any choices, the backward induction stops already at that level.

Why is the intuitive argument in~\cite{Culnane16benalohGT} wrong? After all, if the voter assigns a positive probability $p$ to auditing in the round $\nmax-1$, she knows she will be cheated (in the final round) with exactly that probability. The problem is, if she sets $p = 0$, she is sure to get cheated right away! Thus, the voter should use $p$ to keep the opponent uncertain about her current action, which is the usual purpose of randomizing in strategies.

\subsection{Mixed Nash Equilibria in Finite Benaloh Games}\label{sec:mixed-finite}

We know from Section~\ref{sec:backinduction} that backward induction does \emph{not} eliminate randomized audit strategies in finite Benaloh games.
The next question is: what Nash equilibria do we obtain?
We start with \emph{mixed strategies}, i.e., ones represented by probability distributions
$s_V = [p^V_{1},\cdots,p^V_{\nmax}]$
and $s_D = [p^D_{1},\cdots,p^D_{\infty}]$, where
$p^V_{n}$ is the probability that the voter casts her vote in round $n$, and
$p^D_{n}$ is the probability that the device cheats for the first time in round $n$.

\para{Support sets of Nash strategies.}
First, observe that there are no subgames outside of the main path in the game tree. Thus, all Nash equilibria are subgame perfect.
Moreover, backward induction eliminates the possibility that the device encrypts truthfully in the last round, hence $p^D_{\infty}=0$ in any Nash equilibrium. Consequently, we can represent $s_D$ by $[p^D_{1},\cdots,p^D_{\nmax}]$.

Secondly, all the other probabilities must be nonzero, see the following lemma.\footnote{
  The proofs of the formal results can be found in \short{the extended version of the paper~\cite{ARXIV-VERSION}}\extended{Appendix~\ref{sec:appendix-proofs}}. }
\begin{restatable}{lemmarep}{supportNE}
\label{prop:supportNE}
If $s_V = [p^V_{1},\cdots,p^V_{\nmax}]$ and $s_D = [p^D_{1},\cdots,p^D_{\nmax}]$ form a Nash equilibrium, then
for all $i=V,D$ and $n=1,\dots,\nmax$ we have $p^i_{n} > 0$.
\end{restatable}

\para{Calculating the audit probabilities.}
We compute $p_1^V,\dots,p_{\nmax}^V$ using the standard necessary condition for Nash equilibrium in mixed strategies~\cite[Lemma~33.2]{Osborne94gamet}.
If $(s_V,s_D)$ is a Nash equilibrium with $p^V_n > 0$ and $p^D_n > 0$ for all $n=1,\dots,\nmax$, then the following conditions must hold:
\begin{enumerate}
\item\label{it:svoter} Every deterministic strategy of $V$ obtains the same payoff against $s_D$, in other words:
  $\forall \ncast,\ncast'\in\set{1,\dots,\nmax}\ .\ u_V(\ncast,s_D) = u_V(\ncast',s_D)$
\item\label{it:sdevice} Every deterministic strategy of $D$ obtains the same payoff against $s_V$, in other words:
  $\forall \nfalse,\nfalse'\in\set{1,\dots,\nmax}\ .\ u_D(s_V,\nfalse) = u_D(s_V,\nfalse')$
\end{enumerate}

Consider condition~(\ref{it:sdevice}). Using the payoffs in Figure~\ref{fig:Benaloh-game-NF}, we get:
\begin{restatable}{lemmarep}{NEvoternecessary}
\label{prop:NE-voter-necessary}
If $s_V = [p^V_{1},\cdots,p^V_{\nmax}]$ is a part of Nash equilibrium then
$p_{n+1}^V\ =\ \frac{\dsucc}{\dsucc+\dfail}\ p_{n}^V$
for every $n\in\set{1,\dots,\nmax-1}$.
\end{restatable}

\begin{restatable}{theoremrep}{NEvoter}
\label{prop:NE-voter}
The mixed voting strategy $s_V = [p^V_{1},\cdots,p^V_{\nmax}]$ is a part of Nash equilibrium iff, for every $n\in\set{1,\dots,\nmax}$:
\[
p_{n}^V\ =\ \frac{(1-R)R^{n-1}}{1-R^{\nmax}}, \text{\qquad where\ } R = \frac{\dsucc}{\dsucc+\dfail}.
\]
\end{restatable}

Indeed, the mixed equilibrium strategy $s_V$ provides no \emph{simple} recipe for the voter. This is evident when we consider concrete payoff values.

\begin{example}\label{ex:NE-mixed}
Take $\nmax=5$ and assume $\dsucc = 1, \dfail=4$, i.e., the opponent fears failure four times more than he values success.
Then, $R=0.2$, and hence $s_V = [0.8, 0.16, 0.032, 0.006, 0.001]$ is the unique equilibrium strategy for the voter.
In other words, the voter should cast immediately with probability $0.8$, audit once and cast in round $2$ with probability $0.16$, and so on.
\end{example}

\subsection{Towards Natural Audit Strategies}\label{sec:natural-finite}

So far, we have considered \emph{mixed strategies} for the voter.
That is, the voter draws $\ncast$ before the game according to the probability distribution $s_V$, and then duly follows the outcome of the draw.
An alternative is to use a \emph{behavioral strategy} $b_V = (b_1^V, \dots, b_{\nmax}^V)$, where the voter does a \emph{fresh} Bernoulli-style lottery with probability of success $b_n^V$ in each subsequent round. If successful, she casts her vote; otherwise, she audits and proceeds to the next round.

\para{Behavioral Nash equilibria.}
First, we observe that the game in Figure~\ref{fig:Benaloh-game-EF} is a game of \emph{perfect recall}, i.e., the players remember all their past observations (in our case, the outcomes of all the previous rounds). Thus, by Kuhn's theorem, mixed and behavioral strategies are outcome-equivalent. In other words, the same outcomes can be obtained if the players randomize before the game or throughout the game.
Below, we characterize the behavioral strategy that corresponds to the mixed strategy of Theorem~\ref{prop:NE-voter}.

\begin{restatable}{theoremrep}{NEvoterbehavioral}
\label{prop:NE-voter-behavioral}
The behavioral voting strategy $b_V = [b^V_{1},\cdots,b^V_{\nmax}]$ is a part of Nash equilibrium iff, for every $n\in\set{1,\dots,\nmax}$:
\[
b_{n}^V\ =\ \frac{1-R}{1-R^{\nmax-n+1}}, \text{\qquad where\ } R = \frac{\dsucc}{\dsucc+\dfail}.
\]
\end{restatable}

\begin{example}\label{ex:NE-behavioral}
The behavioral strategy implementing $s_V\extended{ = [0.8, 0.16, 0.032, 0.006, 0.001]}$ of Example~\ref{ex:NE-mixed} is $b_V = [0.8, 0.801, 0.81, 0.83, 1]$.
That is, the voter casts immediately with probability $0.8$, else audits, randomizes again, and casts with probability $0.801$, and so on.
\end{example}

\para{Behavioral audit strategies are reasonably simple.}
At the first glance, the above behavioral strategy seems difficult to execute, too. We cannot expect the voter to randomize with probability \emph{exactly} $0.8$, then \emph{exactly} $0.801$, etc.
On the other hand, $b_V$ can be approximated reasonably well by the following recipe: ``in each round before $\nmax$, cast with probability close to $0.8$, otherwise audit, randomize freshly, and repeat; in the last round, cast with probability $1$.''
This can be generalized due to the following observation.

In Benaloh games, we can usually assume that $\dfail \gg \dsucc$.
First of all, it is important to realize that the opponent of the voter is not the encrypting device, but a human or organizational perpetrator represented by the device. To be more precise, the strategies in the game are defined by the capabilities of the device, but the incentives are those of the perpetrator. Thus, the utility values defined by $u_D$ should not be read as ``the payoffs of the device,'' but rather the utilities of the external party who rigged the device in order to achieve some political, social, or economic goals.
Secondly, the scope of the opponent's activity is not limited to the interaction with a single voter and to corrupting a single encryption device. Presumably, they must have tampered with multiple devices in order to influence the outcome of the vote.
Consequently, the opponent is in serious trouble if even few devices are caught cheating. This is likely to attract attention and trigger investigation, which may lead to an audit of all the encryption devices, revision or voiding of the votes collected from those that turned out corrupt, and even an arrest and prosecution of the perpetrator.
All in all, the penalty for fraud detection ($\dfail$) is usually much higher than the reward for a successful swap of a single vote ($\dsucc$).

\begin{restatable}{theoremrep}{behavioralapprox}
\label{prop:behavioral-approx}
If $\frac{\dsucc}{\dfail} \rightarrow 0$, then the equilibrium strategy $b_V$ of the voter converges to the following behavioral strategy:
\[
\widehat{b_{n}^V}\ =\ \left\{\begin{array}{c@{\quad}l}
      \frac{\dfail}{\dsucc+\dfail} & \text{for } n<\nmax \\
      1 & \text{for } n=\nmax
  \end{array}\right.
\]
\end{restatable}

The finite Bernoulli strategy to audit with probability $R = \frac{\dfail}{\dsucc+\dfail}$ in each round except last seems reasonably simple.
By Theorem~\ref{prop:behavioral-approx}, it is also reasonably close to the unique Nash equilibrium.

\begin{figure}[t]
\resizebox{\textwidth}{!}{\begin{tabular}{|c|c|c|}
\hline
$\ncast$ {\large$\backslash$} $\nfalse$ & \cellcolor{yellow!20} $\mathbf{1}$     & \cellcolor{yellow!20} $\mathbf{2}$
\\ \hline
\cellcolor{yellow!20} $\mathbf{1}$  & $-\vfail,\ \devcolor{\dsucc}$ & $\vsucc,\ \devcolor{0}$
\\ \hline
\cellcolor{yellow!20} $\mathbf{2}$ & $-\caudit,\ \devcolor{-\dfail}$ & $-\caudit - \vfail,\ \devcolor{\dsucc}$
\\ \hline
\end{tabular}
\qquad
\begin{tabular}{|c|c|c|}
\hline
{\small $\ncast$} {\large$\backslash$} {\small $\nfalse$} & \cellcolor{yellow!20} $\mathbf{1}$     & \cellcolor{yellow!20} $\mathbf{2}$
\\ \hline
\cellcolor{yellow!20} $\mathbf{1}$  & $-3,\ \devcolor{1}$ & $2,\ \devcolor{0}$
\\ \hline
\cellcolor{yellow!20} $\mathbf{2}$ & $-1,\ \devcolor{-4}$ & $-4,\ \devcolor{1}$
\\ \hline
\end{tabular}
}
\caption{Benaloh game for $\nmax=2$:\ (a) parameterized payoff table;\ (b) concrete payoff table for the values of Example~\ref{ex:NE-simple-payoffs}}
\label{fig:Benaloh-game-NF-simple}
\end{figure}

\para{Making things even simpler for the voter.}
In order to make Benaloh challenge even easier to use, the voting authority can set $\nmax$ accordingly. In particular, it can fix $\nmax=2$, i.e., allow the voter to audit at most once. That does not seem very restrictive, as empirical evidence suggests that voters seldom audit their votes~\cite{Weber09usabilityHelios,Acemyan14usabilityE3EVV,Ehin22votingEstonia}, and even fewer are able to complete it correctly~\cite{Weber09usabilityHelios,Acemyan14usabilityE3EVV,Gjosteen16experimentNorway}.\footnote{
  In fairness, there is also some evidence that suggests the contrary~\cite[Section~5.6.1]{Gjosteen16evotingNorway}. }
\nocite{Marky18reallyVote}
The Benaloh game in strategic form for $\nmax=2$ is shown in Figure~\ref{fig:Benaloh-game-NF-simple}a.

\begin{restatable}{theoremrep}{behavioralsimple}
\label{prop:behavioral-simple}
For $\nmax = 2$, the behavioral NE strategy of the voter is:
\[
b_{1}^V\ =\ \frac{\dsucc+\dfail}{2\dsucc+\dfail},\qquad\qquad
b_{2}^V\ =\ 1 .
\]
\end{restatable}

To make the analysis intuitive, consider the concrete values in Example~\ref{ex:NE-mixed}.

\begin{example}\label{ex:NE-simple}
Take $\dsucc = 1, \dfail=4$.
By Theorem~\ref{prop:NE-voter-behavioral}, the behavioral Nash equilibrium strategy of the voter is $b_V = [\frac{5}{6}, 1]$.
That is, the voter casts immediately with probability $\frac{5}{6}$, otherwise audits and casts in the next round -- which is a rather simple strategy.
\end{example}

Also, recall our argument that, typically, $\dfail \gg \dsucc$.
In that case, $p_V^1$ becomes close to $1$. In other words, the voter should \emph{almost always} cast immediately, which is a very simple recipe to follow.
Thus, contrary to what Culnane and Teague claim in~\cite{Culnane16benalohGT}, Benaloh challenge can be designed in a way that admits simple Nash equilibrium strategies of the voter.

\subsection{Behavioral Audit Strategies are Simple Enough, But Are They Good Enough?}

We have just seen that finite Benaloh games do allow for simple and easy to use Nash equilibrium strategies.
This seems good news, but what kind of utility do they promise for the voter? That is, how much will the voter benefit from playing NE in Benaloh challenge?
For easier reading, we calculate the answer on our running example.

\begin{example}\label{ex:NE-simple-payoffs}
Following Example~\ref{ex:NE-simple}, we take $\nmax=2, \dsucc = 1, \dfail=4$.
Moreover, we assume $\vsucc=2, \vfail=3, \caudit=1$, i.e., the voter loses slightly more by getting cheated than she gains by casting successfully, and the cost of an audit is half of the gain from a successful vote. The resulting payoff table is presented in Figure~\ref{fig:Benaloh-game-NF-simple}b.

We can now compute the Nash equilibrium strategy of the device using Lemma~\ref{prop:supportNE} and Condition~\ref{it:svoter} of Section~\ref{sec:mixed-finite}.
Consequently, we get
$-3 p_1^D + 2 (1-p_1^D) = -p_1^D -4(1-p_1^D)$, and thus $s_D = [\frac{3}{4}, \frac{1}{4}]$.
Recall that the NE strategy of the voter is $s_V = [\frac{5}{6}, \frac{1}{6}]$.
This yields the following expected payoffs of the players:
\begin{small}
\begin{eqnarray*}
u_V(s_V,s_D) &=&
  -3\frac{15}{24} + 2\frac{5}{24} - 1\frac{3}{24} - 4\frac{1}{24} =\
  -\frac{7}{6}
\\
u_D(s_V,s_D) &=&
  1\frac{15}{24} + 0\frac{5}{24} - 4\frac{3}{24} + \frac{1}{24} =\
  \frac{1}{6}\ .
\end{eqnarray*}
\end{small}
\end{example}

So, the voter gets negative expected utility, and would be better off by not joining the game at all!
If that is the case, then a considerate election authority should forbid electronic voting \emph{not} because there are no simple NE strategies to audit and vote, but because there is one and it is bad for the voter.
The big question is: does Nash equilibrium really provide the right solution concept for rational interaction in Benaloh challenge? We discuss this in Section~\ref{sec:Benaloh-Stackelberg}.

\section{Benaloh According to Stackelberg}\label{sec:Benaloh-Stackelberg}

Nash equilibrium encodes a particular view of rational decision making.
In this section, we discuss its applicability to Benaloh games, suggest that Stackelberg equilibrium is a much better match, and analyze Benaloh challenge through the lens of Stackelberg games.

\subsection{Game-Theoretic Intermezzo, Part Two}\label{sec:GT-primer-short-2}

Every solution concept encodes its own assumptions about the nature of interaction between players and their deliberation processes.
The assumptions behind Nash equilibrium in 2-player games can be characterized as follows~\cite{Perea07doxasticNash}:
\begin{enumerate2}
\item Alice and Bob have common belief that each of them plays best response to one another, and
\item Alice believes that Bob has an accurate view of her beliefs, and that Bob believes that Alice has an accurate view of his beliefs,
\item ...and analogously for Bob.
\end{enumerate2}
Alternatively, NE can be characterized as a local optimum of strategy search with mutual adaptations.
Informally, it represents collective behaviors that can emerge when the agents play the game repeatedly, and adapt their choices to what they expect from the other agents.
Thus, it captures the ``organic'' emergence of behavior through a sequence of strategy adjustments that leads to a point where nobody is tempted to change their strategy anymore.

Is Nash equilibrium the right concept of rationality for Benaloh games? Note that the characterizations of NE are inherently symmetric. In particular, they assume that both players are able to form accurate beliefs about each other's intentions.
This is \emph{not} the case in Benaloh challenge. In line with the arguments of~\cite{Culnane16benalohGT}, the perpetrator has significant technological and motivational advantage over an average voter. For example, he can use opinion polls and statistical methods to get a good view of the voter's preferences.
Even more importantly, machine learning techniques can be used to profile the frequencies with which the voter chooses to audit or cast.
On the other hand, the voter has neither data nor resources to form accurate predictions w.r.t. the strategy of the encryption device.
This seems pretty close to the Stackelberg model of economic interaction.

\para{Stackelberg equilibrium.}
\emph{Stackelberg games}~\short{\cite{Stackelberg52market}}\extended{\cite{Stackelberg34equilibrium,Stackelberg52market}} represent interaction where the strategy of one player (called the \emph{leader}) is known in advance by the other player (the \emph{follower}).
The follower is assumed to play best response to that strategy.
The \emph{generalized Stackelberg equilibrium (SE)}~\cite{Leitmann78stackelberg} prescribes the leader's strategy that maximizes the guaranteed payoff against the follower's best responses.
We define and analyze SE for Benaloh games in Section~\ref{sec:stackelberg-strats}.

\subsection{Pretty Good Strategies against Best Response}\label{sec:stackelberg-strats}

For simplicity, we assume that $\nmax=2$ throughout this section, i.e., the voter can audit the encryption at most once.
Thus, the strategy of the voter can be represented by the probability $p_V$ of casting the vote in the first round.
Similarly, the strategy of the device can be represented by the probability $p^D$ of cheating in the first round.
We first establish $D$'s best response to any fixed $p^V$ and the voter's guaranteed expected utility against best response.
These can be formally defined as follows.

\begin{definition}
The \emph{best response} of $D$, given $V$'s strategy represented by $p^V$, returns those strategies $p^D$ for which the expected value of $u_D(p^V,p^D)$ is maximal:
\[
BR_D(p^V) = \argmax_{p^D\in[0,1]} (E u_D(p^V,p^D)).
\]
\end{definition}
Note that a best response always exists, though it does not have to be unique.

\begin{definition}
The \emph{generalized Stackelberg equilibrium} for $V$ is defined as the strategy that maximizes $V$'s expected payoff against best response.
In case of multiple best responses to some $p^V$, we look at the worst case scenario.
\[
SE_V = \argmax_{p^V\in[0,1]} \inf{}_{p^D\in BR_D(p^V)} (E u_V(p^V,p^D)).
\]
\end{definition}

For randomized strategies of the leader, the Stackelberg equilibrium does not have to exist (cf.~Example~\ref{ex:stackelberg}).
To characterize the leader's abilities in such games, we propose the notion of \emph{Stackelberg value}.
\begin{definition}
The \emph{Stackelberg value} for $V$ is the expected guaranteed payoff that $V$ can obtain against best response \emph{in the limit}:
\[
\StackVal = \sup{}_{p^V\in[0,1]} \inf{}_{p^D\in BR_D(p^V)} (E u_V(p^V,p^D)).
\]
\end{definition}

Clearly, $\StackVal$ is always well defined.
Moreover, the game has a Stackelberg equilibrium if $V$ obtains the Stackelberg value for some strategy.
Finally, for each $\epsilon>0$, the voter has a strategy that $\epsilon$-approximates the Stackelberg value, i.e., obtains at least $\StackVal-\epsilon$ against best response.

\begin{restatable}{lemmarep}{bestresponse}
\label{prop:best-response}
The best response of the device to any fixed strategy of the voter is
\[
BR_D(p^V) = \left\{ \begin{array}{ll}
      0 & \text{for } p^V < p^V_\mathNE \smallskip \\
      1 & \text{for } p^V > p^V_\mathNE \smallskip \\
      \text{any } p^D\in [0,1]\quad{} & \text{for } p^V = p^V_\mathNE
  \end{array}\right.
\]
where $p^V_\mathNE = \frac{\dsucc+\dfail}{2\dsucc+\dfail}$ is the NE probability of casting in round 1.
\end{restatable}

\begin{restatable}{lemmarep}{utilitybestresponse}
\label{prop:utility-best-response}
The voter's expected utility against best response is:
\[
Eu_V(p^V,BR_D(p^V)) = \left\{ \begin{array}{ll}
      p^V\vsucc - (1-p^V)(\caudit+\vfail) & \text{\ for } p^V < p^V_\mathNE \smallskip \\
      -p^V\vfail - (1-p^V)\caudit & \text{\ for } p^V \ge p^V_\mathNE
  \end{array}\right.
\]
\end{restatable}

\begin{figure}[t]\centering
    \includegraphics[width=8cm]{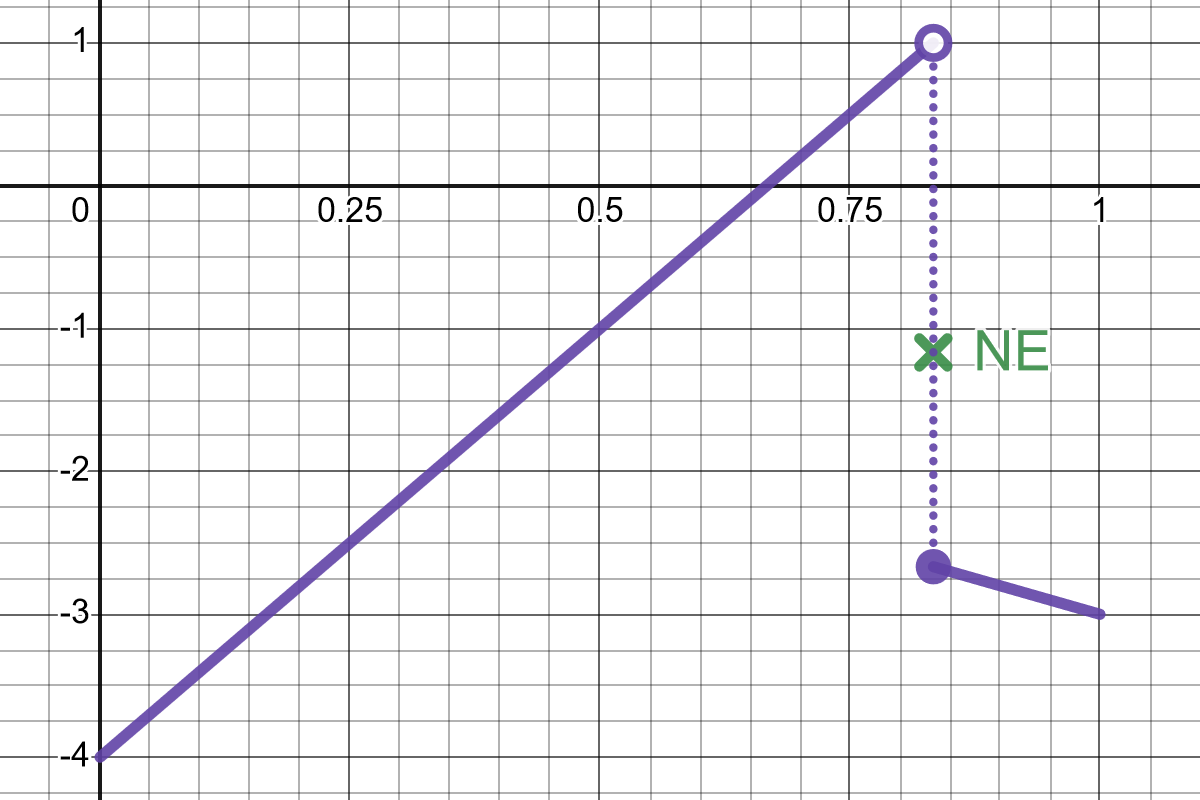}
    \caption{$V$'s payoffs against best response for the Benaloh game in Figure~\ref{fig:Benaloh-game-NF-simple}b. The voter's payoff obtained by Nash equilibrium is shown for comparison}
    \label{fig:stackelberg}
\end{figure}

\begin{example}\label{ex:stackelberg}
The graph of $Eu_V(p^V,BR_D(p^V))$ for the parameters in Example~\ref{ex:NE-simple-payoffs}
(i.e., $\nmax=2, \dsucc = 1, \dfail=4, \vsucc=2, \vfail=3, \caudit=1$)
is depicted in Figure~\ref{fig:stackelberg}.
It is easy to see that the function does not reach its optimum, and hence the optimal $p^V$ against best response does not exist.
Still, the strategies based on $p^V$ being \emph{slightly smaller} than the Nash equilibrium strategy $p^V_\mathNE = \frac{5}{6}$ are quite attractive to the voter, since they obtain payoff that is both positive and strictly higher than the Nash payoff.
\end{example}

The next and final theorem generalizes the example to arbitrary two-round Benaloh games.
It shows that the voter has no optimal Stackelberg strategy in the game (point \ref{it:no-stackelberg}),
but the value of $\StackVal = \frac{\dsucc(\vsucc-\vfail-\caudit) + \dfail\vsucc}{2\dsucc+\dfail}$ can be approximated arbitrarily closely (point \ref{it:stackelberg-value}).
That is, for each $\epsilon>0$, the voter has a strategy that obtains at least $\StackVal - \epsilon$ against best response.
Moreover, $\epsilon$-approximating Stackelberg equilibrium is strictly better than playing Nash equilibrium (point \ref{it:stackelberg-vs-nash}).
Lastly, approximate Stackelberg strategies obtain positive utility for the voter under reasonable assumptions (point \ref{it:stackelberg-positive}).

\begin{restatable}{theoremrep}{stackelberg}
\label{prop:stackelberg}
The following properties hold for the Benaloh game with $\nmax=2$:
\begin{enumerate}
\item\label{it:no-stackelberg}
  There is no Stackelberg equilibrium for $V$ in randomized strategies.
\item\label{it:stackelberg-value}
  The Stackelberg value of the game is
  $\StackVal = \frac{\dsucc(\vsucc-\vfail-\caudit) + \dfail\vsucc}{2\dsucc+\dfail}$.
\item\label{it:stackelberg-vs-nash}
  $\StackVal > Eu_V(p^V_\mathNE,p^D_\mathNE)$, where $(p^V_\mathNE,p^D_\mathNE)$ is the Nash equilibrium.
\item\label{it:stackelberg-positive}
  If $\dfail \gg \dsucc$ and $\vsucc \ge a\vfail$ for a fixed $a>0$, then $\StackVal > 0$.
\end{enumerate}
\end{restatable}

Thus, Stackelberg games capture the rational interaction in Benaloh games better than Nash equilibrium, {and} predict strictly higher payoffs for the voter.

\section{Conclusions, or What Do We Learn from That?}\label{sec:discussion}

In this paper, we analyze a simple game-theoretic model of incentives in Benaloh challenge, inspired by~\cite{Culnane16benalohGT}.
Contrary to~\cite{Culnane16benalohGT}, we conclude that the voters have at their disposal simple strategies to audit and cast their votes.
This is especially the case if encryption audits are limited to at most one audit per voter.
In that event, a pretty good strategy for the voter is to almost always (but not \emph{exactly} always!) cast immediately in the first round.
Interestingly, this is how voters usually behave in real-life elections, according to empirical evidence.

Moreover, we point out that rational interaction in Benaloh games is better captured by Stackelberg equilibrium, rather than Nash equilibrium.
While the optimal Stackelberg strategy is not attainable for the voter, it can be approximated arbitrarily close by casting the vote immediately with probability \emph{slightly lower} than for the Nash equilibrium.
This is good news, because Stackelberg strategies (even approximate) promise strictly better payoffs for the voter than Nash strategies.
And, under reasonable assumptions, they produce positive utility for $V$. Thus, using Benaloh challenge \emph{is} beneficial to the voter, after all.

The takeaway advice based on this study can be summarized as follows:
\begin{enumerate}
\item
  Using Benaloh challenge is practical and beneficial to the rational voter.

\item
  Putting a strict limit on the number of allowed audits makes things easier for the voter.
  The election authority might design the voting system so that each voter can audit the vote encryption at most once.

\item
  The voters should not try to adapt to the strategy of the attacker, the way Nash equilibrium prescribes.
  Instead, they should stick to auditing the votes with a fixed (and rather low) frequency, thus approximating the Stackelberg optimum and putting the opponent on the defensive.
\end{enumerate}

\para{Discussion and future work.}
An obvious limitation of the current study is the assumption of \emph{complete information} about the structure of the game. In particular, it is dubious to assume that the voter knows how much the adversary values the outcomes of the game.
In the future, we plan to extend the analysis to an incomplete information game model of Benaloh challenge, e.g., in the form of a Bayesian game~\cite{Harsanyi72generalized}.

\todo{add short discussion of the payoffs}
Moreover, the analysis in this paper is performed as a 2-player game between a single voter and the voter's device. It would be interesting to see how this extends to scenarios where the adversary controls multiple devices and plays multiple rounds with different voters.
Last but not least, the players' payoffs for either failing or succeeding need further discussion. In particular, we assume that the costs of failure for the opponent are much higher than the benefits of success; this should be better justified or refuted.

\para{Acknowledgments.} The author thanks Stanisław Ambroszkiewicz, Peter B. Roenne, Peter Y.A. Ryan, and the anonymous reviewers of E-VOTE-ID for their valuable comments, suggestions, and discussions.
The work has been supported by NCBR Poland and FNR Luxembourg under the PolLux/FNR-CORE projects STV (POLLUX-VII/1/2019 and C18/IS/12685695/IS/STV/Ryan), SpaceVote (POLLUX-XI/14/SpaceVote/2023 and C22/IS/17232062/SpaceVote) and PABLO (C21/IS/16326754/PABLO).

\bibliographystyle{plain}

\extended{
  \clearpage
  \appendix

\section{Formal Proofs}\label{sec:appendix-proofs}

Here, we present the proofs of our formal results.

\subsection{Proofs of Section~\ref{sec:Benaloh-Nash} (Benaloh According to Nash)}

\smallskip
\supportNE*

\smallskip
\begin{proof}
Suppose that $(s_V,s_D)$ is a Nash equilibrium, and that $p^V_n = 0$ for some $n$ (i.e., the voter always audits in round $n$). Take the smallest such $n$. Then, $s_D = n$ is the unique best response of $D$, i.e., the device must cheat for the first time in that round.
We consider two cases now:
(i) $n=1$: in that case, the voter is better off playing $s_V=1$, i.e., casting deterministically at the first round.
(ii) $n>1$: in that case, the voter is better off by swapping $p_{n-1}^V$ and $p_n^V$, i.e., postponing the action planned for round $n-1$ until round $n$.
In both cases, we get that $(s_V,s_D)$ is not a Nash equilibrium, which is a contradiction.
Hence, we get that $p^V_n > 0$ for all $n$. [*]

Suppose now that $p^D_n = 0$ for some $n$ (i.e., the device never cheats in round $n$). Take the smallest such $n$. If $n=1$, then $V$'s best response is $s_V = 1$, which contradicts [*].
If $n>1$, then $V$'s best response includes $p_{n-1}^V = 0$, i.e., $V$ postpones casting at $n-1$ until the next round, which also contradicts [*].
Hence, also $p^D_n > 0$ for all $n$.
\qed
\end{proof}

\smallskip
\NEvoternecessary*

\smallskip
\begin{proof}
Recall Condition~(\ref{it:sdevice}), saying that:
$$\forall \nfalse,\nfalse'\in\set{1,\dots,\nmax}\ .\ u_D(s_V,\nfalse) = u_D(s_V,\nfalse').$$
It is equivalent to:
\begin{eqnarray*}
\forall n\in\set{1,\dots,\nmax-1}\ &.&\ u_D(s_V,n+1) - u_D(s_V,n) = 0\qquad [*] \end{eqnarray*}
Notice that:
\begin{footnotesize}
\begin{eqnarray*}
u_D(s_V,n) &=& \sum_{i=1}^{\nmax} p_i^V\cdot u_D(i,n) = \\
 &=& \sum_{i=1}^{n-1} p_i^V\cdot 0 + p_{n}^V\cdot\dsucc + \sum_{i=n+1}^{\nmax} p_i^V\cdot(-\vfail) \\
 &=& \dsucc\cdot p_{n}^V - \dfail\cdot \sum_{i=n+1}^{\nmax} p_i^V
\end{eqnarray*}
\end{footnotesize}
Similarly,
\begin{footnotesize}
\begin{eqnarray*}
u_D(s_V,n+1) &=& \sum_{i=1}^{\nmax} p_i^V\cdot u_V(i,n+1) = \\
 &=& \dsucc\cdot p_{n+1}^V - \dfail \cdot\sum_{i=n+2}^{\nmax} p_i^V
\end{eqnarray*}
\end{footnotesize}
By this and [*], we get that:
\begin{footnotesize}
\[
\dsucc\cdot p_{n+1}^V - \dsucc\cdot p_{n}^V + \dfail\cdot p_{n+1}^V\ =\ 0
\]
\end{footnotesize}
In consequence,
\begin{footnotesize}
\[
p_{n+1}^V = \frac{\dsucc}{\dsucc+\dfail} p_{n}^V
\]
\end{footnotesize}
which completes the proof.
\qed
\end{proof}

\smallskip
\NEvoter*

\smallskip
\begin{proof}
If $s_V$ is a part of Nash equilibrium then $p_n^V>0$ for all $n=1,\dots,\nmax$ (by Lemma~\ref{prop:supportNE}).
Moreover, by Lemma~\ref{prop:NE-voter-necessary}, the probabilities $p_1^V,\dots,p_{\nmax}^V$ form a geometric sequence with ratio $R = \frac{\dsucc}{\dsucc+\dfail}$.
Thus, $\sum_{n=1}^\nmax p_n^V = p_1^V\cdot \frac{1-R^\nmax}{1-R}$ must be equal to $1$.
In consequence, $p_1^V = \frac{1-R}{1-R^\nmax}$, and hence $p_{n}^V\ =\ \frac{(1-R)R^{n-1}}{1-R^{\nmax}}$.

Notice that the above probability distribution is the only admissible solution, i.e., no other $s_V$ can be a part of Nash equilibrium.
By Nash's theorem, the finite Benaloh game must have at least one equilibrium; hence, it is the unique one.
\qed
\end{proof}

\smallskip
\NEvoterbehavioral*

\smallskip
\begin{proof}
We claim that the above behavioral strategy implements the unique Nash equilibrium strategy $s_V=[p_1^V,\dots,p_\nmax^V]$ of Theorem~\ref{prop:NE-voter}. To prove this, it suffices to verify that
$p_n^V = (1-b_1^V)\cdot \dots \cdot(1-b_{n-1}^V) \cdot b_n^V$ for all $n=1,\dots,\nmax$.
That is, casting at round $n$ indeed corresponds to unsuccessful Bernoulli trials in the first $n-1$ rounds, and a successful trial in round $n$.
The check is technical but straightforward.
\extended{\todo{Complete, see the handwritten notes!}}
\qed
\end{proof}

\smallskip
\behavioralapprox*

\smallskip
\begin{proof}
Take the behavioral NE strategy $b_V$ in Theorem~\ref{prop:NE-voter-behavioral}.
For $\frac{\dsucc}{\dfail} \rightarrow 0$, we get $R \rightarrow 0$. Hence, $1-R^{\nmax-n+1}$ for $n<\nmax$ converges to $1$ much faster than $1-R$, and thus $b_{n}^V = \frac{1-R}{1-R^{\nmax-n+1}}$ gets arbitrarily close to $1-R \extended{= 1 - \frac{\dsucc}{\dsucc+\dfail}}= \frac{\dfail}{\dsucc+\dfail}$.
\qed
\end{proof}

\smallskip
\behavioralsimple*

\smallskip
\begin{proof}
Fix $\nmax=2$. By Theorem~\ref{prop:NE-voter-behavioral}, we get $b_{1}^V = \frac{1-R}{1-R^2} = \frac{1}{1+R} = \frac{\dsucc+\dfail}{2\dsucc+\dfail}$.
Similarly, $b_{2}^V = \frac{1-R}{1-R} = 1$.
\qed
\end{proof}

\subsection{Proofs of Section~\ref{sec:Benaloh-Stackelberg} (Benaloh According to Stackelberg)}

\smallskip
\bestresponse*

\smallskip
\begin{proof}
Given a strategy profile represented by $(p^V,p^D)$, the expected payoff of the device is:
\begin{eqnarray*}
Eu_D(p^V,p^D) &\ =\ & p^Vp^D\dsucc - (1-p^V)p^D\dfail + (1-p^V)(1-p^D)\dsucc \\
   & = & (2p^V\dsucc + p^V\dfail - \dsucc - \dfail)p^D + (1-p^V)\dsucc .
\end{eqnarray*}
Therefore, the derivative of $Eu_D(p^V,p^D)$ is
\[
\frac{d Eu_D(p^V,p^D)}{d p^D}\ =\ 2p^V\dsucc + p^V\dfail - \dsucc - \dfail ,
\]
which is negative for $p^V < \frac{\dsucc+\dfail}{2\dsucc+\dfail}$ and positive for $p^V > \frac{\dsucc+\dfail}{2\dsucc+\dfail}$.
We recall from Theorem~\ref{prop:behavioral-simple} that $p^V_\mathNE=\frac{\dsucc+\dfail}{2\dsucc+\dfail}$ is the Nash equilibrium probability that the voter casts in the first round.\footnote{
  Note that, for $\nmax=2$, mixed and behavioral strategies coincide and can be used interchangeably. }
Thus, $Eu_D(p^V,p^D)$ is decreasing for $p^D \in [0,p^V_\mathNE)$, and hence reaches its maximum at $p^D = 0$.
Similarly, $Eu_D(p^V,p^D)$ is increasing for $p^D \in (p^V_\mathNE,1]$, and has its maximum at $p^D = 1$.

Finally, by Lemma~\ref{prop:supportNE} and the necessary Nash condition~(\ref{it:sdevice}), any response of $D$ to strategy represented by $p^V_\mathNE$ must obtain the same expected payoff for $D$, hence each is a best response.
\qed
\end{proof}

\smallskip
\utilitybestresponse*

\smallskip
\begin{proof}
For $p^V < p^V_\mathNE$, we have $E u_V(p^V,BR_D(p^V)) = E u_v(p^V,0) = p^V\vsucc - (1-p^V)(\caudit+\vfail)$.
Similarly, for $p^V > p^V_\mathNE$, we have $E u_V(p^V,BR_D(p^V)) = E u_v(p^V,1) = -p^V\vfail - (1-p^V)\caudit$.

For $p^V = p^V_\mathNE$, any $p^D\in[0,1]$ is a best response. Since $Eu_V(p^V_\mathNE,p^D)$ is a linear function w.r.t. $p^D$, it reaches its minimum for either $p^D=0$ or $p^D=1$.
Observe that $Eu_V(p^V_\mathNE,0) - Eu_V(p^V_\mathNE,1) = (2\vfail+\vsucc)p^V_\mathNE - \vfail > 0$
because
$p^V_\mathNE = \frac{\dsucc+\dfail}{2\dsucc+\dfail} > \frac{1}{2} > \frac{\vfail}{2\vfail+\vsucc}$ .
Thus, $Eu_V(p^V_\mathNE,0) > Eu_V(p^V_\mathNE,1)$, and $V$'s lowest payoff against best response at $p^V_\mathNE$ is $Eu_V(p^V_\mathNE,1)$.
\qed
\end{proof}

\smallskip
\stackelberg*

\smallskip
\begin{proof}
\underline{Ad.~\ref{it:no-stackelberg} \&~\ref{it:stackelberg-value}}:
Consider $f(p^v) = Eu_V(p^V,BR_D(p^V))$, established in Lemma~\ref{prop:utility-best-response}. The function is increasing for $p^V\in[0,p^V_\mathNE)$ and decreasing for $p^V\in[p^V_\mathNE,1]$.
Moreover, $\lim_{p^V \rightarrow (p^V_\mathNE)^{^-}} f(p^V) = Eu_V(p^V_\mathNE,0) > Eu_V(p^V_\mathNE,1) = f(p^V_\mathNE))$.
Thus, $\StackVal = \sup_{p^V\in[0,1]} f(p^V) = Eu_V(p^V_\mathNE,0) = p^V_\mathNE\vsucc - (1-p^V_\mathNE)(\caudit+\vfail) = \frac{(\dsucc+\dfail)(\vsucc+\vfail+\caudit)}{2\dsucc+\dfail}$,
and the value is not reached by any $p^V$.
\qed

\smallskip\noindent
\underline{Ad.~\ref{it:stackelberg-vs-nash}}:
By Lemma~\ref{prop:supportNE}, $p^D_\mathNE > 0$.
Moreover, $Eu_V(p^V_\mathNE,p^D)$ is linear w.r.t. $p^D$, and we already know that $Eu_V(p^V_\mathNE,0) > Eu_V(p^V_\mathNE,1)$, thus it must be strictly decreasing.
In consequence, $\StackVal = Eu_V(p^V_\mathNE,0) > Eu_V(p^V_\mathNE,p^D_\mathNE)$.
\qed

\smallskip\noindent
\underline{Ad.~\ref{it:stackelberg-positive}}:
Let $\vsucc \ge a\vfail$, and recall that $\caudit<\vfail$.
Then,
$\StackVal \ge \frac{\dsucc(a\vfail-\vfail-\vfail) + a\dfail\vfail}{2\dsucc+\dfail} = \vfail(a - \frac{(2+a)\dsucc}{2\dsucc+\dfail})$.
For $\frac{\dsucc}{\dfail} \rightarrow 0$, this converges to $a\vfail$, which is greater than $0$.
\qed
\end{proof}

 }

\end{document}